# A far Ultra-Violet Polarimeter by reflection for Pollux on board LUVOIR

Maëlle Le Gal[a], Arturo López Ariste[b], Coralie Neiner[a]

[a]LESIA, Observatoire de Paris, PSL Research University, CNRS, Sorbonne Université, Univ. Paris Diderot, Sorbonne Paris Cité, 5 place Jules Janssen, 92195 Meudon, France; [b]Institut de Recherche en Astrophysique et Planétologie, CNRS, IRAP, France

## ABSTRACT

The ultra-violet (UV) high-resolution spectropolarimeter Pollux is being studied in Europe under CNES leadership for the LUVOIR space mission. LUVOIR is a projected 15-m telescope equipped with a suite of instruments proposed to NASA. Pollux will perform spectropolarimetric measurements from 90 to 400 nm with a resolution of 120000. The spectrograph will be divided in three channels, each with its own polarimeter: far UV (FUV, 90-124.5 nm), mid UV (MUV, 118.5-195 nm), and near UV (NUV, 190-390 nm). We present here our FUV prototype and our investigation to optimize this polarimeter (angle, materials, coating…).

## 1. INTRODUCTION

Although some spectropolarimeters have been developed for a short wavelength UV range (e.g. for $Ly_\alpha$), none has ever been built to cover dozens to hundreds of nm at once. We present here our work for the design of a FUV polarimeter for Pollux on board LUVOIR. Basic polarimeters are made of a modulator and an analyzer, usually working by transmission and birefringence. However, working at such a short wavelength is a real challenge as we do not know any birefringent material that transmits light under 120 nm. Therefore, we studied an innovative reflection-based polarimeter for the FUV. Based on the theory of reflection polarimeters [1], we studied the modulation and polarimetric efficiency of a prototype consisting of a 3-reflection modulator, called a K-mirror, and a reflection analyzer. We chose to use 3 mirrors as it is the smallest number of mirrors that permits not to deviate the beam from the optical axis and have an optimal intensity. At each reflection, the beam goes through a phase shift between p and s polarization depending on the incident angle on the mirror. By rotating the three mirrors as a block, we can create a temporal modulation. The analyzer is a reflection on a plate based on the effect of a Brewster-like angle to select the polarization we want to measure. The principle of this new type of polarimeter is shown in figure 1. The selection of the material for the mirrors and their coating is also a challenge as optical indices are not well known at these wavelengths. A R&T study on UV coatings funded by CNES and performed by REOSC is ongoing to study this issue.

## 2. SIMULATION OF THE POLARIMETER

### 2.1 Mueller matrix

**Modulator**

The modulator is made of three mirrors which rotate as a block. Each reflexion impacts the beam, represented by the Stokes vector and can be modelled by the reflexion Mueller matrix:

$$M_R(i,\lambda) = \begin{pmatrix} X(i,\lambda)^2 + 1 & X(i,\lambda)^2 - 1 & 0 & 0 \\ X(i,\lambda)^2 - 1 & X(i,\lambda)^2 + 1 & 0 & 0 \\ 0 & 0 & 2X(i,\lambda)cos(\tau(i,\lambda)) & 2X(i,\lambda)sin(\tau(i,\lambda)) \\ 0 & 0 & -2X(i,\lambda)sin(\tau(i,\lambda)) & 2X(i,\lambda)cos(\tau(i,\lambda)) \end{pmatrix}$$

with $X^2 = r_\parallel^2/r_\perp^2$ the ratio of Fresnel amplitude reflexion coefficients and with $\tau$ the difference of phase shift between both polarizations. This matrix is obviously dependent on the angle of incidence i and the wavelength $\lambda$ but also on the optical index of the mirror, the one of the coating and its thickness. The Mueller matrix of the modulator is then $M_{modulator} = M_R(i,\lambda) * M_R(2i - \frac{\pi}{2},\lambda) * M_R(i,\lambda)$. When modulating, the rotation is given by the rotation matrix: $R(\theta) = \begin{pmatrix} 1 & 0 & 0 & 0 \\ 0 & cos(2\theta) & sin(2\theta) & 0 \\ 0 & -sin(2\theta) & cos(2\theta) & 0 \\ 0 & 0 & 0 & 1 \end{pmatrix}$.

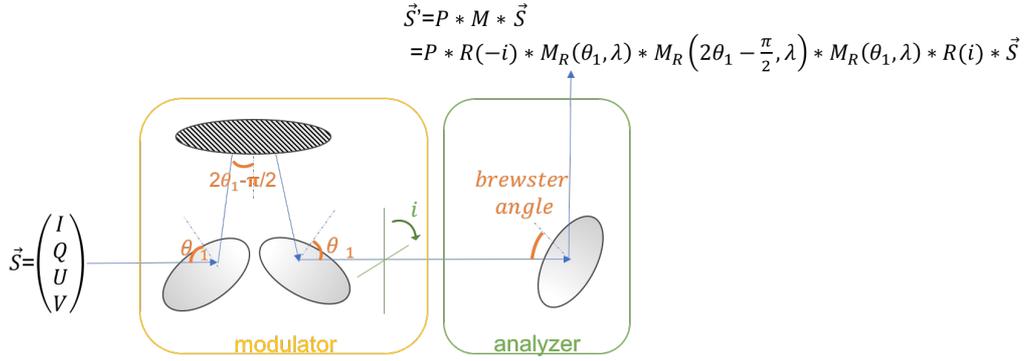

Figure 1: Schematic view of a polarimeter using only mirrors.

**Analyzer**

To polarize the light, we do not have any other choice than to use a mirror at the Brewster angle. This creates a non-negligible issue: a chromatic effect. Indeed, the Brewster angle depends on the wavelength and we cannot have a 100% efficient polarizer at all wavelengths. The Mueller matrix of the analyzer is then a reflexion matrix.

**Global Matrix and (de)modulation matrix**

The global Mueller matrix of the polarimeter is then:

$$M = M_{analyzer}(\beta) * R(-\theta) * M_{modulator} * R(\theta)$$
$$= M_{analyzer}(\beta) * R(-\theta) * M_R(i,\lambda) * M_R(2i - \frac{\pi}{2},\lambda) * M_R(i,\lambda) * R(\theta)$$

Unfortunately, we can measure only intensities, which are given by the first line of the matrix. We introduce the **Modulation Matrix** which is the first line of the global matrix for each modulation angle. The modulation matrix answers the equation: $I_{out} = M_{modulation} * S_{in}$, with $I_{out}$ the matrix of the measurement made for each angle of modulation and $S_{in}$ the Stokes vector we want to measure. To retrieve the Stokes vector, we have to used the demodulation matrix, which is the pseudo-inverse of the modulation matrix: $M_{demodulation} = (M_{modulation}^T \cdot M_{modulation})^{-1} \cdot M_{modulation}^T$. We then have $S_{in} = M_{demodulation} * I_{out}$, to retrieve directly the initial Stokes vector from the intensity measurements. The demodulation matrix will be used to determine the efficiency of the polarimeter, see section 3.2.1.

## 2.1 Reflectivity and phase shift

### 2.1.1 Without coating

Here we present the modulation of a K-mirror made of a metal without coating, we use the following equations based on [2]:

$$\chi^2 = \frac{f^2+g^2-2f*sin(\theta)*tan(\theta)+sin^2(\theta)*tan^2(\theta)}{f^2+g^2+2f*sin(\theta)*tan(\theta)+sin^2(\theta)*tan^2(\theta)} \text{ and } tan(\tau) = \frac{2g*sin(\theta)*tan(\theta)}{sin^2(\theta)*tan^2(\theta)-(f^2+g^2)}$$

with $\theta$ the angle of incidence and with

$$f^2 = \frac{1}{2}(n^2 - k^2 - sin^2(\theta) + \sqrt{(n^2 - k^2 - sin^2(\theta))^2 + 4n^2k^2}$$

and

$$g^2 = \frac{1}{2}(k^2 - n^2 + sin^2(\theta) + \sqrt{(n^2 - k^2 - sin^2(\theta))^2 + 4n^2k^2}.$$

### 2.1.2 With coating

Here we consider the addition of a coating on the three mirrors.

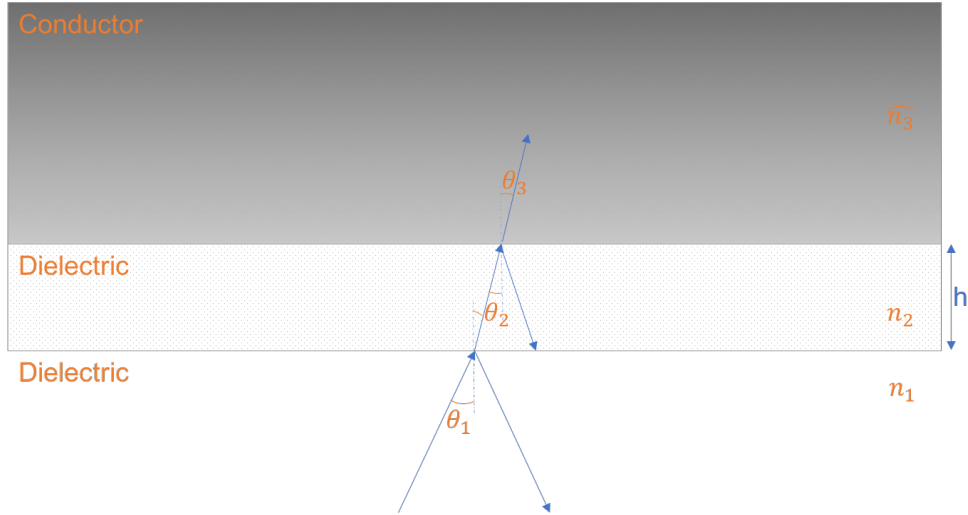

Figure 2: Reflection on a surface with a coating.

To determine $\chi$ and $\tau$ as introduced in section 2.1, we need to calculate the reflectivity and the phase shift for each polarization ($\parallel$ and $\perp$). In [3], the reflexion with a transparent film on an absorbing substrate is given by:

$$r = \frac{r_{12}+\rho_{23}e^{i(\phi_{23}+2\beta)}}{1+r_{12}\rho_{23}e^{i(\phi_{23}+2\beta)}} \text{ and } tan(\delta_r) = \frac{\rho_{23}(1-r_{12}^2)sin(\phi_{23}+2\beta)}{r_{12}(1+\rho_{23}^2)+\rho_{23}(1+r_{12}^2)cos(\phi_{23}+2\beta)}$$

with $\widehat{n_2}cos(\theta_2) = u_2 + iv_2$, $\widehat{n_3}cos(\theta_3) = u_3 + iv_3$ and $\beta = 2\pi n_2 \frac{h}{\lambda} cos(\theta_2)$

$$r_{12,\perp} = \frac{n_1 cos(\theta_1) - u_2 + iv_2}{n_1 cos(\theta_1) + u_2 + iv_2}$$

$$\rho_{23,\perp} = \frac{(n_2 cos(\theta_2) - u_3)^2 + v_3^2}{(n_2 cos(\theta_2) + u_3)^2 + v_3^2}$$

$$tan(\phi_{23,\perp}) = \frac{2v_3 n_2 cos(\theta_2)}{u_3^2 + v_3^2 - n_2^2 cos(\theta_2)^2}$$

and

$$r_{12,\|} = \frac{cos(\theta_1)/n_1 - cos(\theta_2)/\widehat{n_2}}{cos(\theta_1)/n_1 + cos(\theta_2)/\widehat{n_2}}$$

$$\rho_{23,\|} = \frac{[n_3^2(1-\kappa_3^2)cos(\theta_2) - n_2 u_3]^2 + [2n_3^2 \kappa_3 cos(\theta_2) - n_2 v_3]^2}{[n_3^2(1-\kappa_3^2)cos(\theta_2) + n_2 u_3]^2 + [2n_3^2 \kappa_3 cos(\theta_2) + n_2 v_3]^2}$$

$$tan(\phi_{23,\|}) = 2n_2 n_3^2 cos(\theta_2) \frac{2\kappa_3 u_3 - (1-\kappa_3^2)v_3}{n_3^4(1+\kappa_3^2)^2 cos(\theta_2)^2 - n_1^2(u_3^2+v_3^2)^2}.$$

As shown in figure 2, we call 1 the environment of the mirror, vacuum in our case. We call 2 the coating and 3 the mirror. To model the polarimeter, we need complex indices of the materials we want to use. Little can be found in the literature, therefore we set up our own experiment to measure them.

## 3. EXPERIMENT TO RETRIEVE COMPLEX INDICES OF MATERIALS

### 3.1 Principle of the experiment

To optimize this polarimeter, complex indices have to be well known at every studied wavelength. Indeed, the previous equations in section 2.1 show that the parameters of the modulation of the polarimeter are complex indices, wavelength, and incidence angles. At the moment, SiC and ta-C, a DLC material, are being considered for Pollux. Unfortunately, the indices found in the literature are not coherent between different sources. To measure indices at short wavelengths, it has thus been decided to set up a gold reflective polarimeter. The polarimeter has the same design as the one studied above, and is composed of pure gold. The use of gold allows us to know precisely what to expect as it is a well-known material. Moreover, the parameters of gold are constant with the supplier and process. The reflectivity is low, but since this is only for tests, it is not important and is compensated by long exposure times. The principle of measurement is the same as the polarimeter described above. The gold modulator rotates around the optical axis and modulates the light while a fourth mirror at the analog of the Brewster angle polarizes the light. The studied material sample is placed at the entrance of the gold polarimeter, and we measure the Stokes vector after reflection on the sample for several incidence angles. Having the Stokes vector at the output of the source and the one after reflection for several different angles allows us to retrieve the sample's complex index.

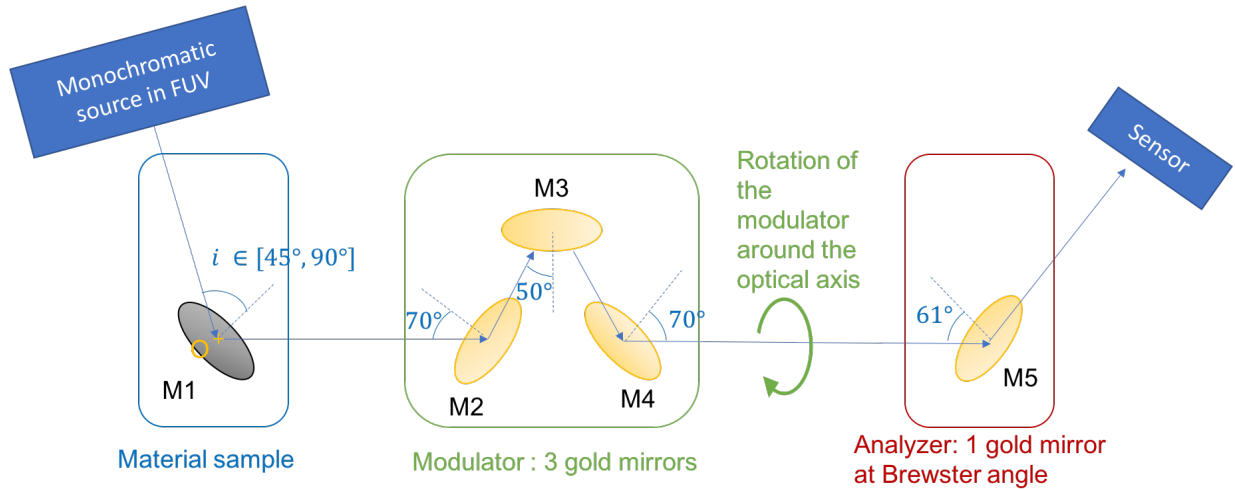

Figure 3: Experiment using a gold polarimeter in order to retrieve the complex indices of samples of material

## 3.2 Simulation

### 3.2.1 Modulation of the gold polarimeter

The modulation is given by the same expressions as in section 2.1.1 as we are in the case of pure gold without coating. For example for a 70° incidence and 4 modulation angles (30°, 74°, 105° and 149°), we obtain the modulation shown in figure 4.

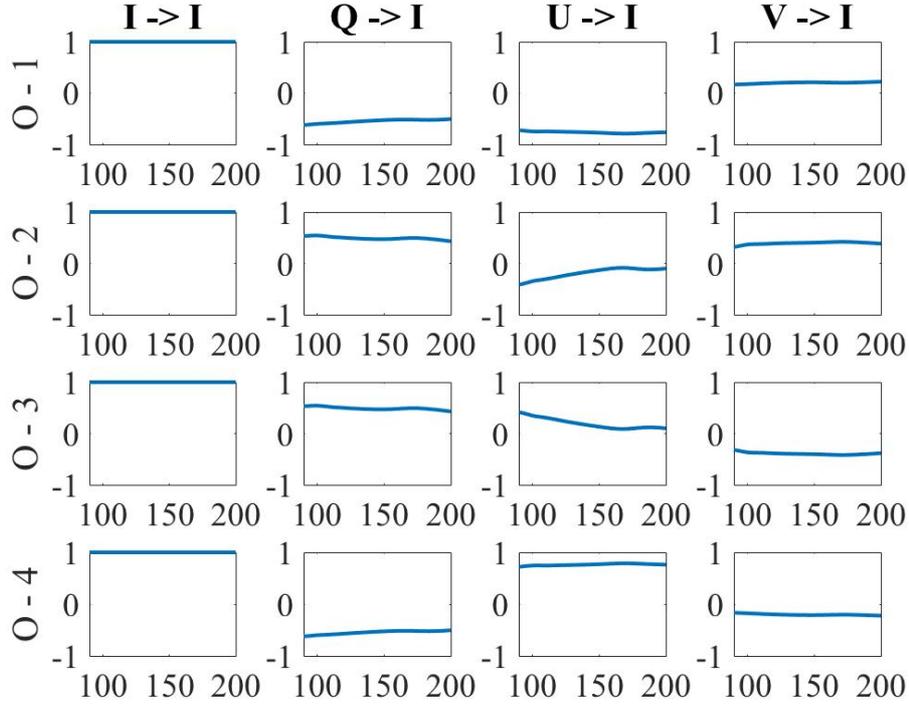

Figure 4: Modulation by a gold K-mirror rotating at 4 angles with an incidence of 70°.

**Efficiency**

To determine the efficiency of the modulation, we introduce polarimetric efficiency based on the demodulation matrix [5]:

$$\epsilon_i = (n \sum_{j=1}^{n} M^2_{demodulation,ij})^{-\frac{1}{2}}$$ with i $\in$ [I,Q,U,V], n the number of measurements (number of modulation angles).

To achieve the determination of all the Stokes parameters simultaneously, the optimal efficiency is is $\frac{1}{\sqrt{3}}$ (because $\epsilon_Q^2 + \epsilon_U^2 + \epsilon_V^2 \leq 1$). For this experiment, the coefficients can be measured independently which improves the efficiencies. For example for a 70° incidence and 4 modulation angles (30°, 74°, 105° and 149°) and simultaneous measurements, the polarimetric efficiencies are shown in figure 5. The efficiencies are quite low for this example because of the gold. The optimal material for the FUV domain, which we will use for Pollux, will improve the efficiencies and the reflectivity.

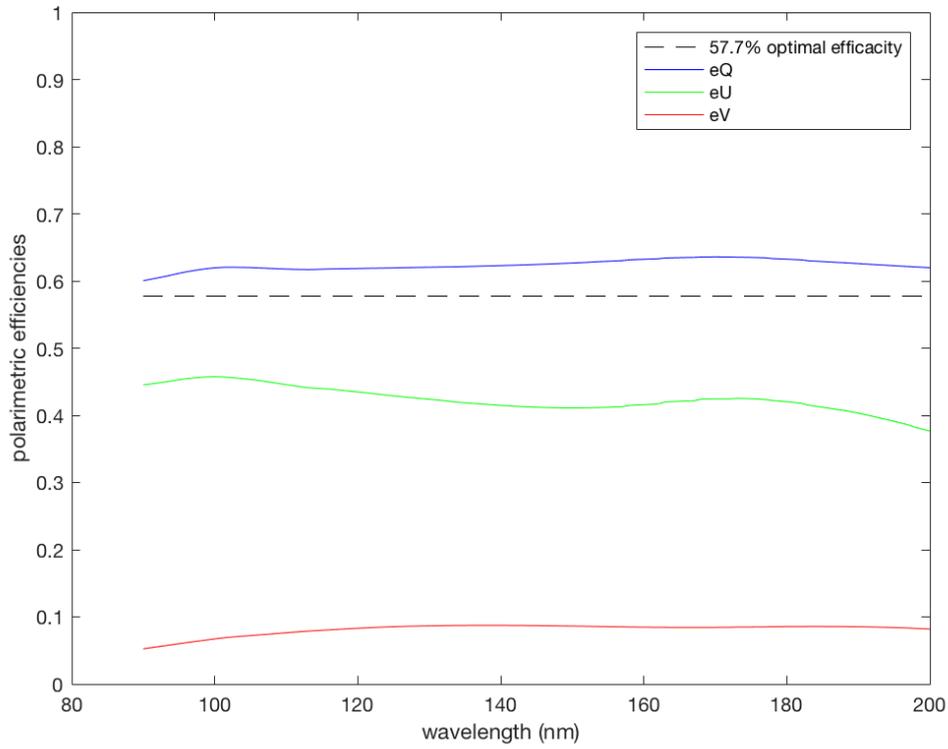

Figure 5: Polarimetric efficiencies for a gold polarimeter with 4 angles of modulation (30°, 74°, 105° and 149°).

**3.2.2 Analyzer**

To polarize the light, as explained in section 2.0.1, the Brewster angle is being considered. We are looking for the metal equivalent that has the best contrast between s and p polarization for gold. The contrast is defined as $C = \frac{R_s - R_p}{R_s + R_p}$. We consider wavelengths from 90 to 135 nm. In figure 6, by displaying the contrast as a function of the angle of incidence and as a function of wavelength, we can find the incidence for which the contrast is maximal for most wavelengths. We can also see that the contrast is more or less constant with wavelength, which means the chromatic effect is negligible. By averaging the contrast results over wavelength, a maximum is easily spotted at $i_{max} = 61°$. The mean contrast is then 70%.

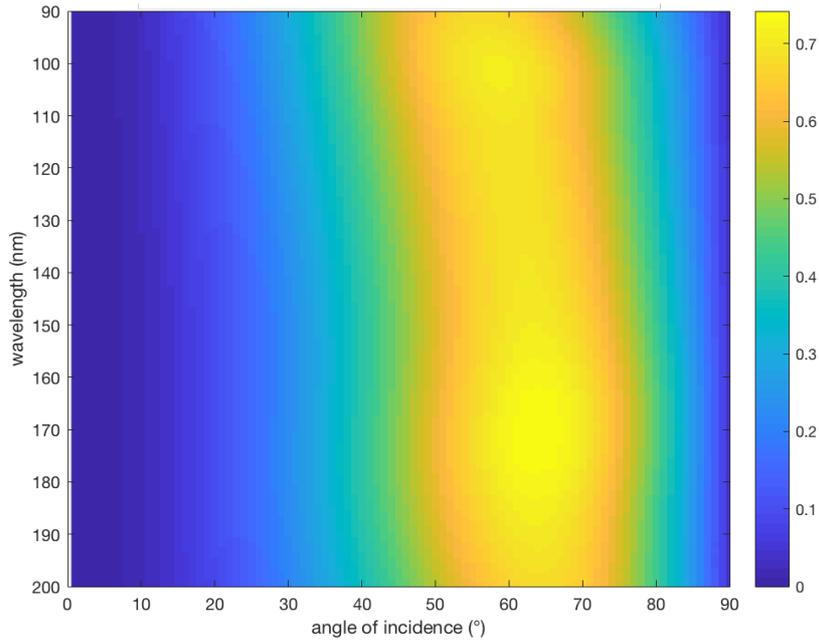

Figure 6: Contrast between s and p light as a function of incidence angle (in abscissa, from 0° to 90°) and as a function of wavelength (in ordinate, from 90 to 200 nm) for a gold mirror.

To evaluate the global efficiency of the polarizer, we introduce the figure of merit $\varepsilon = C\sqrt{\frac{R_s+R_p}{2}}$. $\varepsilon$ is maximum for a perfect polarizer and equals 0.7071. In figure 7, we show the figure of merit as a function of wavelength. The mean of the figure of merit is 0.3610. This means that our polarizer is not perfect and does not have a 100% efficiency. We have to take this into account for the measurements.

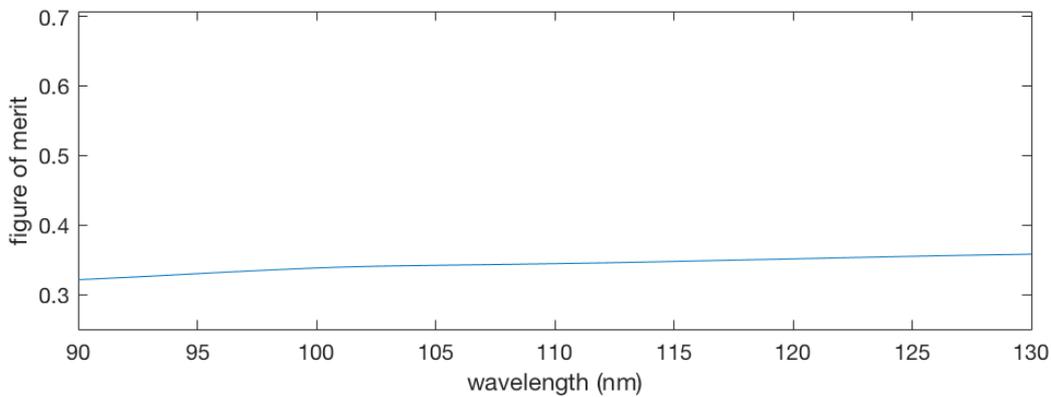

Figure 7: Figure of merit of the gold Brewster-like analyzer (gold mirror at 61°) as a function of wavelength.

## 4. CONCLUSION

A new type of polarimeter has been studied to perform polarimetry in the FUV domain. With or without coating, we can calculate the performances (modulation, polarimetric efficiencies and efficiency of the analyzer) of this polarimeter. In order to design the best FUV polarimeter, an experiment has been set up to measure the polarimetric properties of the materials considered for Pollux, using a gold polarimeter. The study of this gold polarimeter constitutes an example of what we can build for the Pollux FUV channel but its performances will be better once it is optimized with the right material.

**Acknowledgments**

The authors would like to thank Jean-François Mariscal and the LATMOS team for hosting this experiment and sharing their knowledge, Napoléon Nguyen Tuong and Jean-Philippe Amans for their mechanical expertise, as well as Safran REOSC for the loan of the material samples for the experiment.